\documentclass[12pt,preprint]{aastex}
\shorttitle{25 Ori}
\shortauthors{McGehee}

\begin{document}

\title{The Southern Flanking Fields of the 25 Orionis Group}


\author{Peregrine M. McGehee{\altaffilmark{\ref{LANL1}}}}

\altaffiltext{1} {
Los Alamos National Laboratory,
LANSCE-IC, 
MS H820,
Los Alamos, NM 87545; \\
peregrine@lanl.gov
\label{LANL1}}


\begin{abstract}
The stellar group surrounding the Be (B1Vpe) star 
25 Orionis was discovered to be a pre-main-sequence population
by the Centro de Investigaciones de Astronomia (CIDA) Orion
Variability Survey and subsequent spectroscopy. We analyze Sloan
Digital Sky Survey multi-epoch photometry to map the southern extent of
the 25 Ori group and to characterize its pre-main-sequence population.
We compare this group to the neighboring Orion
OB1a and OB1b subassociations and to active star formation sites
(NGC 2068/NGC 2071) within the Lynds 1630 dark cloud.
We find that the 25 Ori group has a radius of 1.4$\degr$, corresponding
to 8-11 pc at the distances of Orion OB1a and OB1b.
Given that the characteristic sizes of young
open clusters are a few pc or less this suggests that 25 Ori
is an unbound association rather than an open cluster.
Due to its PMS population having a low Classical
T Tauri fraction ($\sim$10\%) we conclude that
the 25 Ori group is of comparable age to the 11 Myr Orion OB1a 
subassociation.
\end{abstract}



\section{Introduction}

The pre-main-sequence 25 Orionis group has been recently
identified by the Centro 
de Investigaciones 
de Astronomia Variability Survey of Orion \citep[CVSO;][]{bri05a}
on the basis of multi-epoch imaging and follow-up
spectroscopy.
25 Ori has also
been noted by \citet{kha05} as a candidate open cluster, ASCC 16,
based on
analysis of the All-Sky Compiled Catalogue of 2.5 Million Stars 
\citep[ASCC-2.5;][]{kha01}. This group is named after the
Be star 25 Orionis, $(\alpha_{2000}, \delta_{2000})$ = (05$^h$24$^m$44.92$^s$, 
+01$\degr$50$^m$47.2$^s$) or $(l,b)$ = (200.96, -18.2).
Spectroscopically confirmed candidate group members are found within 
one degree from 25 Orionis \citep{bri05b}.

We use multi-epoch imaging by the Sloan Digital Sky Survey (SDSS)
to select candidate pre-main-sequence (PMS) objects based on variability
and location in color-magnitude diagrams.
The SDSS imaging within the Orion region spans
the   
declination range of -1.25$\degr$ to +1.25$\degr$, thereby
covering
the southern edge of the 25 Ori group within 0.6$\degr$
of the central Be star.

In this paper we examine the PMS populations near 25 Orionis to
determine the observational characteristics and spatial extent
of this group. In \S 2 we describe the SDSS equatorial imaging 
survey in the Orion region. \S 3 revisits the classification scheme
of \citet{mcg05} for PMS candidates based on multi-band variability.
In \S 4 we present
our results, and in \S 5 we discuss our findings, conclusions, and
future work.

\section{Observations}

\subsection{Photometry}

The SDSS low Galactic latitude data which includes the Orion equatorial 
imaging used in this work are presented by \citet{fin04}. 
The SDSS multi-epoch imaging data used in this paper as well as
the selection for candidate low-mass
pre-main sequence objects are discussed in \citet{mcg05}.
Here we describe the salient
features of the SDSS drift-scanned imaging.

The SDSS obtains deep 
photometry with asinh magnitude \citep{lup99} limits 
(defined by 95\% detection repeatability 
for point sources) of 
$u=22.0$, $g=22.2$, $r=22.2$, $i=21.2$ and $z=20.5$. These five passbands,
$ugriz$, 
have effective wavelengths of 3540, 4760, 6290, 7690, and
9250 {\AA}, respectively.
A technical summary of the SDSS is given by \citet{yor00}.
The SDSS imaging camera and telescope are described by \citet{gun98}
and \citet{gun06}, respectively. 
\citet{ive04} discuss the data management and photometric quality 
assessment system.

The Early Data Release and the Data Release One are described by
\citet{sto02} and \citet{aba03}. The former includes an extensive 
discussion of the data outputs and software. \citet{pie03} describe the 
astrometric calibration of the survey and the 
network of primary photometric standard stars is described by 
\citet{smi02}. The photometric system itself is defined by \citet{fuk96},
and the system which monitors the site photometricity by
\citet{hog01}.
\citet{aba03} discuss the differences between the native SDSS 2.5m
$ugriz$ system and the $u'g'r'i'z'$ standard star system defined
on the USNO 1.0 m \citep{smi02}.
The Two-Micron All Sky Survey 2MASS \citep[2MASS;][]{skr97} obtained nearly
complete coverage of the sky in $JHK_s$.

\subsection{Comparison of CIDA and SDSS Survey Coverage}

The SDSS Orion equatorial data consist of a 2.5$\degr$ wide stripe
centered on the celestial equator. For this work we consider the
37.5 deg$^2$ bounded in R.A. by
5$^h$ to 6$^h$ that lies completely within the CVSO survey region. 
While the temporal sampling by the SDSS imaging is much sparser
with at most 7 observations per target over a span of nearly 5 years,
the SDSS survey is deeper and thus is able to detect low-mass
and very-low-mass objects.

There are 197 spectroscopically confirmed Weak-lined T Tauris (WTTS) and 
Classical T Tauris (CTTS) found by the CVSO survey \citep{bri05a}. 
Out of the 76 that
are located within the SDSS equatorial scans, we find that 10 
are identified in the
SDSS low-mass star study \citep{mcg05} which selected for M dwarf 
colored stars on the basis of $(r-i) > 0.6$
and $Q_{riz} > 0.35$, where the latter is a reddening-invariant index
formed by the SDSS $riz$ magnitudes (see \S 3.1 below). 
These matches are listed in Table \ref{ori-tbl1}.
The complete set of WTTS and CTTS candidates identified in the SDSS
survey are presented in Table \ref{ori-tbl2}. These tables include
object classification based on multi-band variability into 
non-PMS, WTTS, and CTTS as discussed in \S3.2 below. Within
\S3.3 we investigate the effectiveness of selection and classification
of PMS objects based solely on variability.

\begin{figure}
\plotone{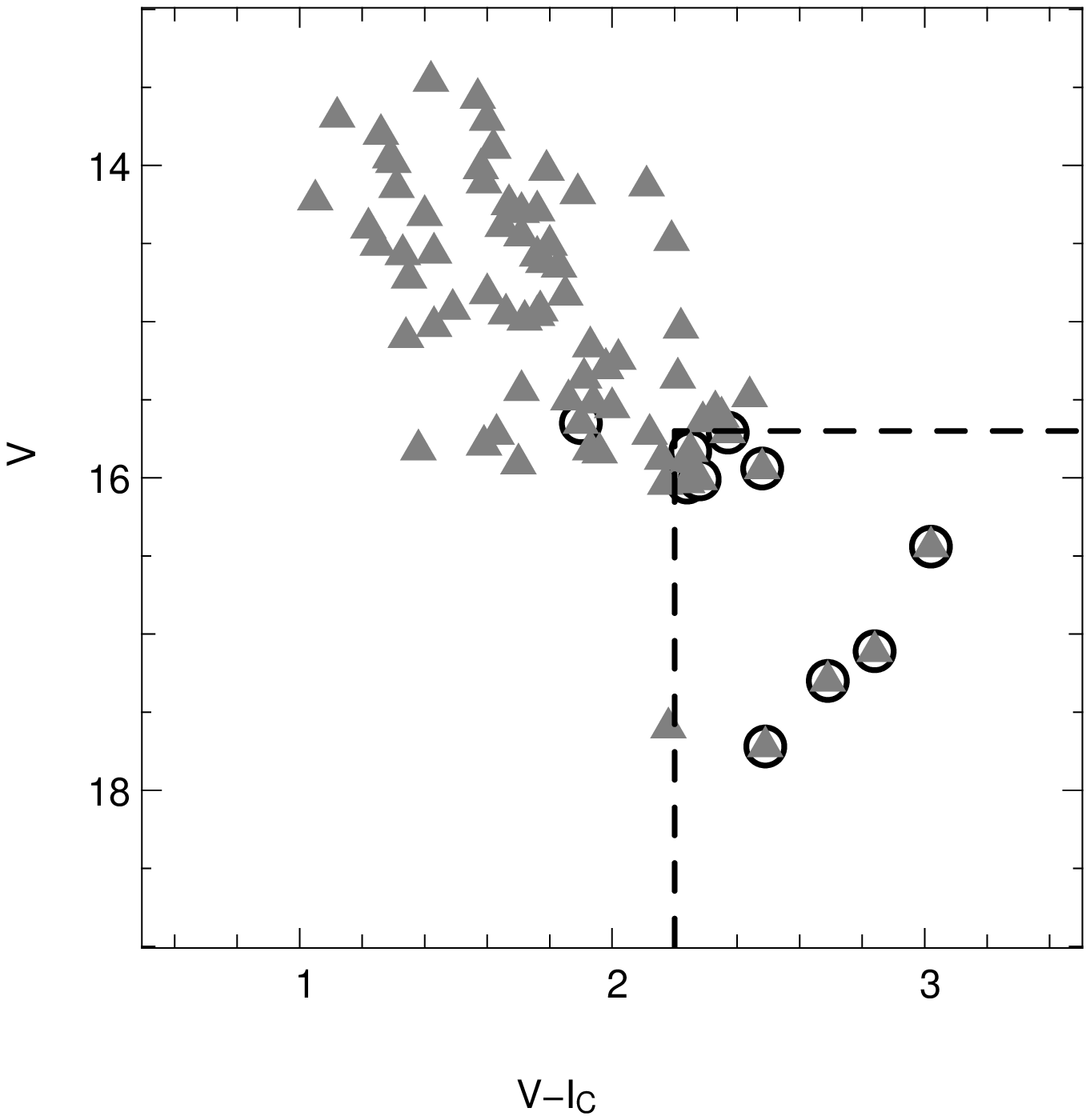}
\caption{{\bf $(V-I_c,V)$ color-magnitude diagram for equatorial CIDA 
objects.} The colors and magnitudes of the 76 spectroscopically confirmed
pre-main-sequence objects found in the CIDA Orion Variability Survey
that lie within the SDSS equatorial scans are shown here. Those objects
selected by the SDSS-based low-mass star study of \citet{mcg05}
are highlighted by open circles. The effective selection region of the
SDSS study is bounded by $V-I_C > 2.2$ and $V > 15.7$ ({\it dashed lines}). 
The faint star at $V-I_C = 2.18, V = 17.60$
is the continuum CTTS CVSO 157. 
%%The BCAH98 isochrones are shown for
%%ages of 1 Myr, 10 Myr, 100 Myr and masses of 0.25 M$_\odot$, 0.50 M$_\odot$,
%%0.75 M$_\odot$, and 1.00 M$_\odot$. The assumed distance is that of the
%%Orion OB1b subassociation (440 pc).
\label{fig-cidavvi}}
\end{figure}

The objects detected by the SDSS low-mass star survey are both faint 
($V > 15.7$) and red ($V-I_C > 2.2$), as evident in 
Figure \ref{fig-cidavvi}. 
The single faint object that
was not selected by \citet{mcg05}
is CVSO 157, a highly veiled continuum CTTS, whose $r-i$ and $Q_{riz}$ colors
were in the ranges 0.42 to 0.77 and -0.50 to -0.11, respectively.
%%In Figure \ref{fig-cidavvi} we have overlaid the \citet{bar98} [BCAH98] 
%%isochrones, assuming that the stars are at the 440 pc distance of the
%%Orion OB1b subassociation.

\section{Identification of CTTS Candidates}

\subsection{Reddening-Invariant Indices}

The Orion OB1a and OB1b subassociations comprise
a range of star formation environments ranging from older,
dispersed populations having low line-of-sight interstellar extinction
to very young and embedded systems, such as those associated with
the Lynds 1630 dark cloud \citep{lyn62}. In order 
to mitigate the need to deredden the photometry for each object
we utilize a number
of reddening-invariant indices (colors) and magnitudes to aid in
the classfication of the stars in these regions 

In this work we employ reddening-invariant indices 
of the form
$Q_{xyz} = (x-y) - (y-z) \times E(x-y)/E(y-z)$.
For passbands defined at wavelengths less than 1$\mu$m, $Q_{xyz}$ 
is dependent upon the assumed ratio of general to selective
extinction ($R_V = A_V/E(B-V)$; \citet{car89}). 
Here $xyz$ refer to the specific 
passbands, e.g.
$ugrizJHK_S$, and $E(x-y)$ is the color excess due to reddening of the
$x-y$ color. This definition of reddening-invariant colors
follows \citet{joh53} in that in our notation their original 
$Q$ would be written as 
$Q_{UBV}$. In the $(x-y,y-z)$ color-color diagram the 
$Q_{xyz}$ axis is perpendicular to the reddening vector. 
These indices are discussed in full by \citet{mcg05}. 

We use the extinction tables derived by 
D. Finkbeiner\footnote{private communication; 
see http://www.astro.princeton.edu/$\sim$dfink/sdssfilters/} 
to define the coefficients used in defining reddening-invariant
colors. These tables contain the $A_X/E(B-V)$ values for the
SDSS $ugriz$ filters for specific values of $R_V$ and source
spectra. The values we present here are obtained using
an F dwarf source spectrum and $R_V$ = 3.1 
which is the standard extinction law found in the diffuse ISM. 

In this work
we use optical and near-IR indices $Q_{riz} = (r-i) - 0.987 (i-z)$
and $Q_{JHK} = (J-H) - 1.563 (H-K)$.
We also employ reddening-invariant magnitudes constructed from
linear combinations of observed magnitudes and colors.
Here we use a near-IR reddening-invariant magnitude
$J_{J-H} = J - (J-H) A_J/E(J-H) = J - 2.76 (J-H)$. The numerical
terms in the $Q_{JHK}$ and $J_{J-H}$ definitions are based
on the reddening coefficients of \citet{sch98}. 
The notation follows that of \citet{bab05}
who used $K_{sJ-K_s}$ to determine distances to red clump
stars in the inner Galactic bulge. We chose the combination
of the $J$ and $H$ bands to minimize the effect of thermal
emission from the inner circumstellar disk.

\subsection{Comparison of Variability and Near-IR Disk Indicators}

In \citet{mcg05} we identified candidate PMS objects on the basis
of $\sigma_g > 0.05$, where the subset flagged as possible
CTTS rather than WTTS were selected by $\sigma_z > 0.05$. 
Here $\sigma_g$ and $\sigma_z$ are the standard deviations in the
$g$ and $z$ bands measured over all observations for a star.
Application of these criteria to a comparison field yielded
possible contamination fractions of 8.7\% for all PMS
candidates (WTTS and CTTS) and of 2.9\% for CTTS only.
The majority of the contaminants are likely foreground active
M dwarfs (dMe).

In this work we improve these selection criteria by comparing the
SDSS multi-band variability against the intrinsic near-IR 
excess computed from 2MASS observations. This procedure assumes
that the high-amplitude variability and near-IR excess are
both due to the presence of an inner circumstellar disk with the
former generated by magnetospheric
accretion processes and the latter by disk thermal emission.

\begin{figure}
\plotone{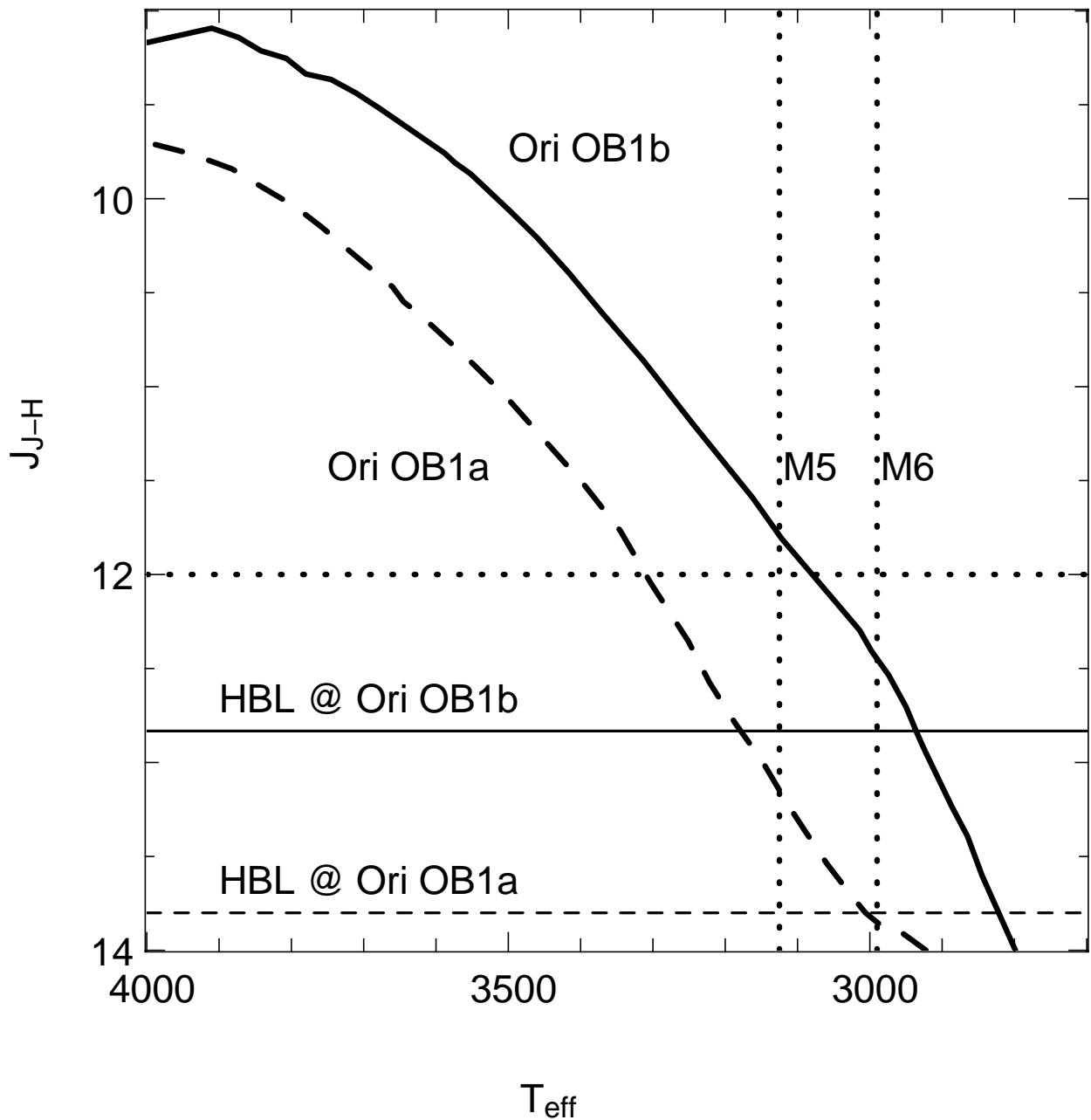}
\caption{{\bf $(J_{J-H}, T_{eff})$ diagram for Orion OB1a and OB1b.}
The relation between $T_{eff}$ and $J_{J-H}$ for the Orion OB1a
({\it dashed line}) and Orion OB1a ({\it solid line}) subassociations
are shown here based on the BCAH98 models. The two vertical dotted
lines mark spectral types of M5 and M6, assuming the PMS temperature
scale of \citet{luh03}. The faint magnitude limit of $J_{J-H} = 12.0$
for detection of circumstellar disk signatures, based on a lack
of contrast in the NIR for spectral types of M6 or later, is traced
by the horizontal dotted line. The Hydrogen Burning Limit (0.075 $M_\odot$) is
indicated for both subassociations.
\label{fig-jjhteff}}
\end{figure}

Due to the low-mass and very-low-mass nature of our sample we
need to be concerned about the contrast between the stellar 
photospheric emission and that from the inner circumstellar
disk. \citet{liu03} note that the $K$ band excess vanishes
for spectral types of M6 and later, thereby making use of longer
wavelength bands such as $L$ and $M$ critical for detection
of circumstellar disks around very-low-mass stars and young
brown dwarfs. In Figure \ref{fig-jjhteff} we see that,
based on the \citet{bar98} [BCAH98] isochrones and the \citet{luh03} PMS
temperature scale, a 2 Myr old star in the Orion OB1b
subassociation with a spectral type between M5 and M6 would
have $J_{J-H} = 12.0$. We adopt this as the faint magnitude
limit for reliable detection of $K$ band disk emission 
in our study.

\begin{figure}
\plotone{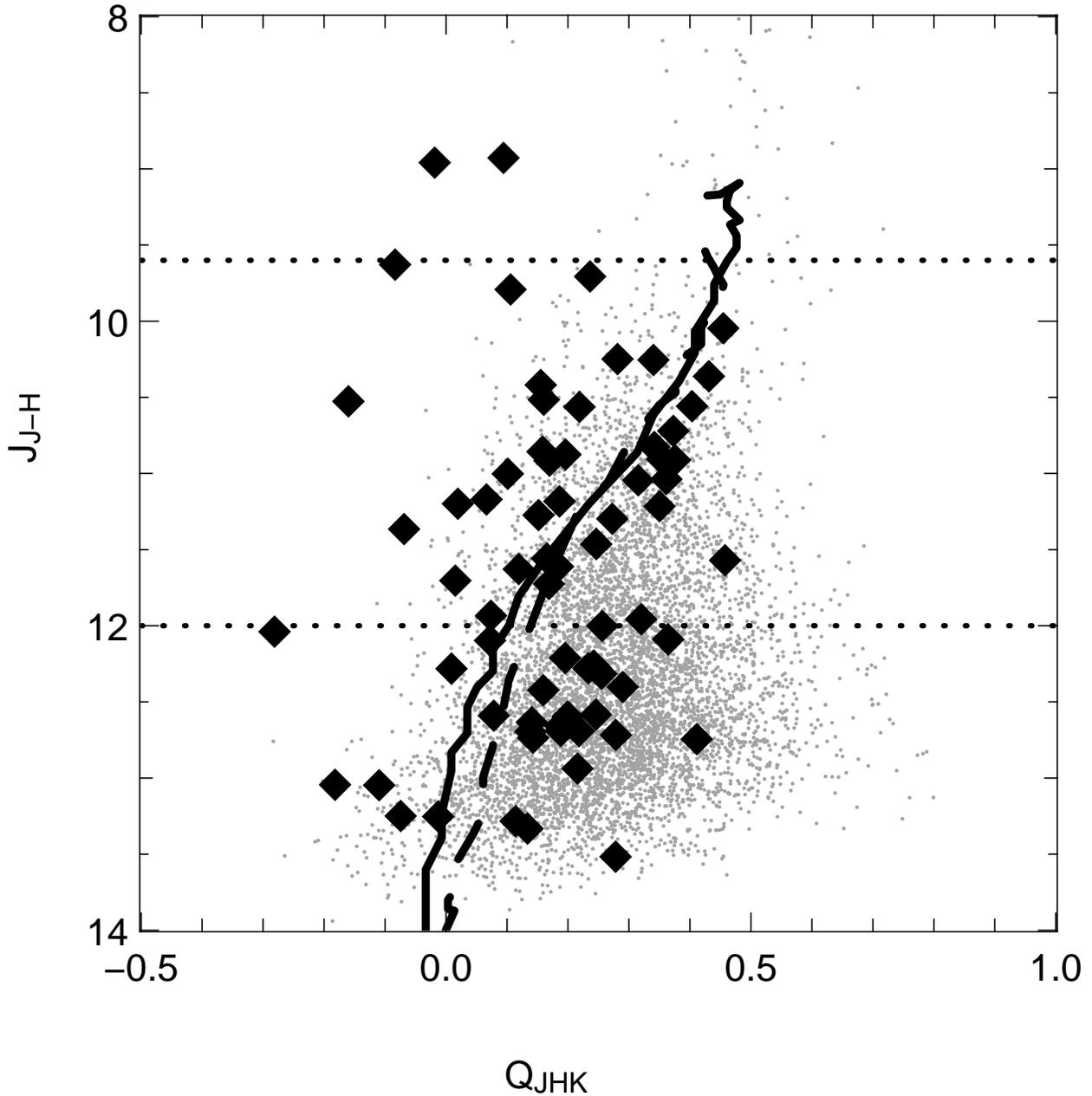}
\caption{{\bf Near-IR color-magnitude diagram for Orion.}
The reddening-invariant $(Q_{JHK}, J_{J-H})$ color-magnitude diagram
for the Orion sample is plotted here with the BCAH98 isochrones for
Orion OB1a ({\it dashed line}) and Orion OB1b ({\it solid line})
shown as references. The reddening-invariant near-IR excess,
${\Delta}Q_{JHK}$, is measured relative to the isochrones for
stars having $9.6 < J_{J-H} < 12.0$ ({\it dotted lines}).
Disk candidates $({\Delta}Q_{JHK} < 0)$ are found to the left
of the isochrones. The CTTS candidates, as inferred by SDSS multi-band
variability (see Figures \ref{fig-dqsigg} and \ref{fig-dqsigz}),
are highlighted by diamonds.
\label{fig-qccd}}
\end{figure}

We begin by defining a reddening-invariant near-IR excess,
${\Delta}Q_{JHK}$, relative to the stellar locus in the
$(Q_{JHK}, J_{J-H})$ color-magnitude diagram.
In Figure \ref{fig-qccd} we present the BCAH98 isochrones
for Orion OB1a and Orion OB1b in this reddening-invariant
near-IR color-magnitude diagram. The steepening of
the stellar locus for the brighter stars motivates us to
define a bright limit of $J_{J-H} = 9.6$. The isochrones for
Orion OB1a and OB1b lie nearly on top of each other for
$9.6 < J_{J-H} < 12.0$. A least-squares fit gives
$Q_{JHK} = 1.986 - 0.157 J_{J-H}$ in this magnitude range, thus
we define ${\Delta}Q_{JHK} = Q_{JHK} - (1.986 - 0.157 J_{J-H})$.

In order to determine the reddening-invariant colors of 
disk candidates we first consider
the CTTS locus of \citet{mey97} which, in the $(J-H,H-K)$ 
color-color diagram, is $(J-H) = 0.58(H-K) + 0.52$. For a star
on the CTTS locus an increase in $(H-K)$ of ${\delta}(H-K)$
implies ${\delta}Q_{JHK} = -0.98 {\delta}(H-K)$ and
${\delta}J_{J-H} = 0.58 {\delta}(H-K)$ with the resulting
${\delta}{\Delta}Q_{JHK} = -1.23 {\delta}(H-K)$. Therefore we
expect CTTS candidates to have ${\Delta}Q_{JHK} < 0$, that is,
found to the left of the stellar locus in the $(Q_{JHK},J_{J-H})$ 
diagram.

\begin{figure}
\plotone{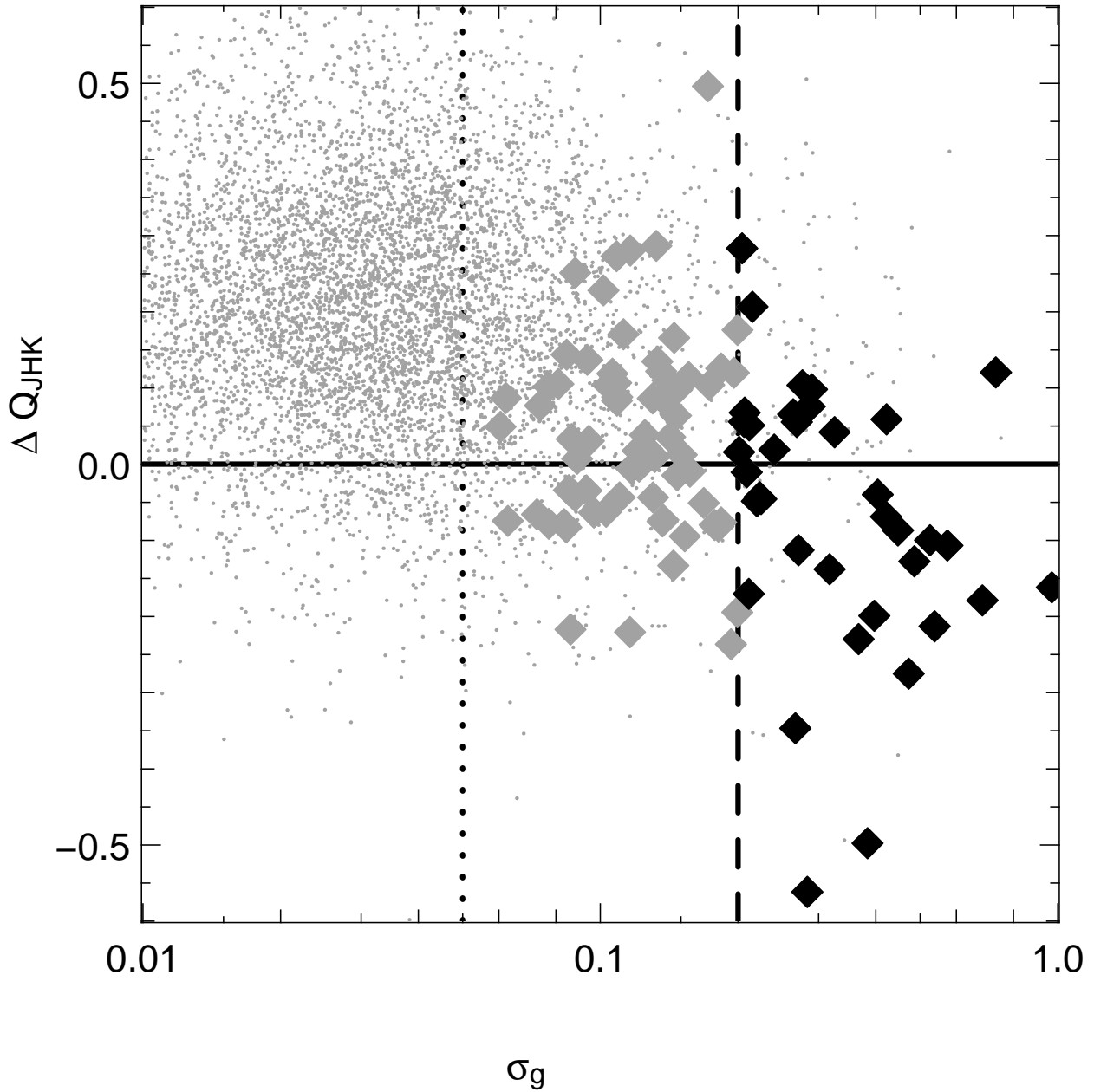}
\caption{{\bf Near-IR excess and $g$ band variability.}
In this figure we compare the intrinsic near-IR excess,
as inferred by ${\Delta}Q_{JHK}$, and $\sigma_g$, the standard deviation
in the $g$ band measured over all observations of a star.
Stars meeting the variability thresholds of \citet{mcg05}
for CTTS candidates ($\sigma_g > 0.05$ ({\it dotted line})
and $\sigma_z > 0.05$) having $9.6 < J_{J-H} < 12.0$ (see
Figure \ref{fig-qccd}) are shown as diamonds. In this work
we increase the $\sigma_g$ threshold to 0.2 magnitudes
({\it dashed line}) due to the marked increase in near-IR excess
at this variability level and above. The stars meeting the new
variability criteria of
$\sigma_g > 0.2$ and $\sigma_z > 0.05$ are shown as
black diamonds.
\label{fig-dqsigg}}
\end{figure}

\begin{figure}
\plotone{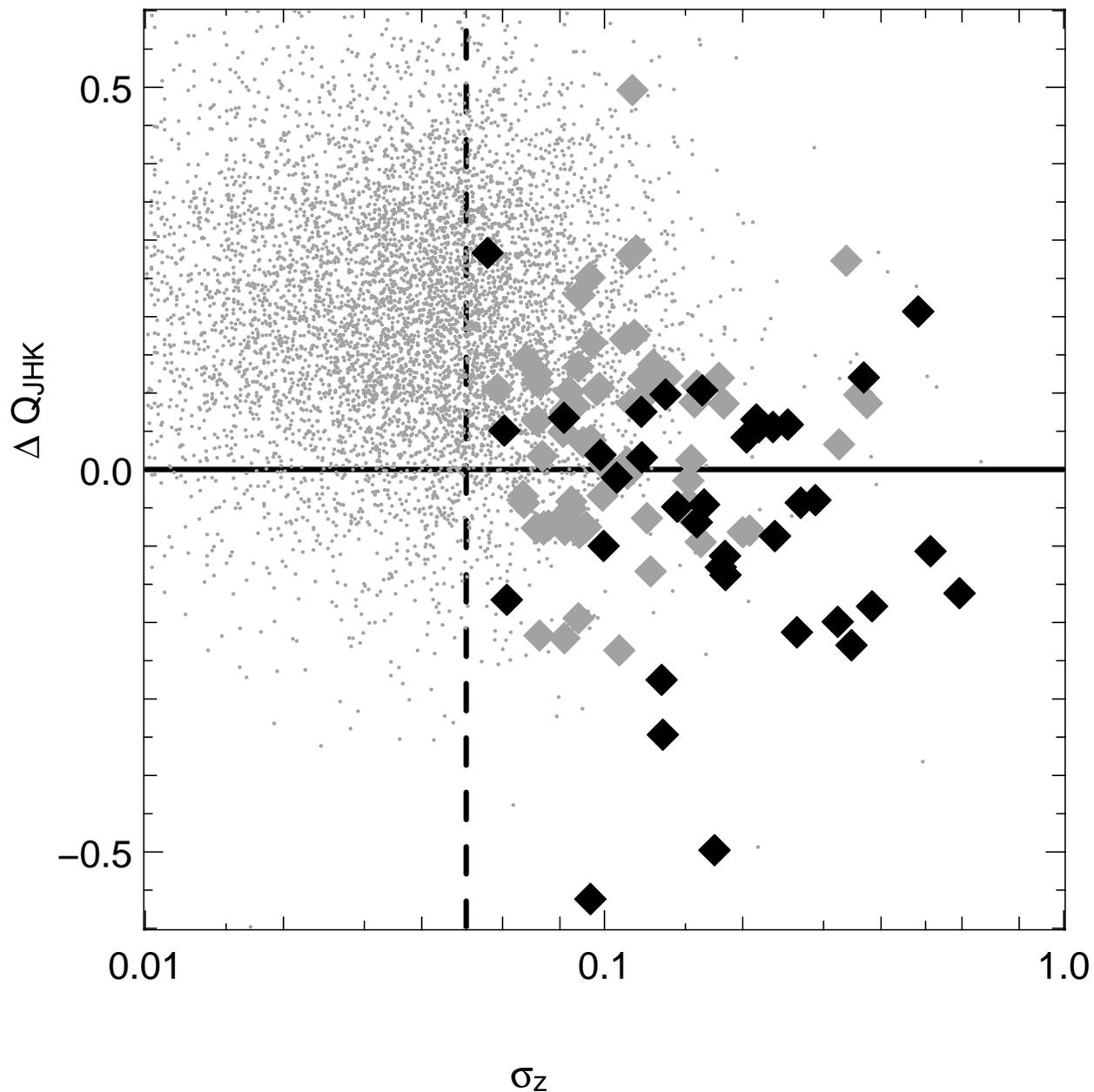}
\caption{{\bf Near-IR excess and $z$ band variability.}
Following Figure \ref{fig-dqsigg}, we compare the 
intrinsic near-IR excess,
as indicated by ${\Delta}Q_{JHK}$, and $\sigma_z$, the standard deviation
in the $z$ band measured over all observations of a star.
The $\sigma_z = 0.05$ threshold used by \citet{mcg05} and this
work is marked by the dashed line. As in Figure \ref{fig-dqsigg}
the diamonds indicate CTTS candidates based on the
criteria in \citet{mcg05} ({\it grey}) and this work
({\it black}).
\label{fig-dqsigz}}
\end{figure}

Examination of Figures \ref{fig-dqsigg} and \ref{fig-dqsigz},
in which the standard deviations in $g$ and $z$ are compared
against ${\Delta}Q_{JHK}$ for stars having $9.6 < J_{J-H} < 12.0$,
reveals that the criteria used by \citet{mcg05} for detection
of CTTS candidates are insufficently stringent. Specifically, the
threshold for variability in the $g$ band that matches the
onset of significant near-IR excess is $\sigma_g = 0.2$, i.e.,
a factor of four greater than used in \citet{mcg05}. The stars
selected by these new criteria ({\it diamonds}) and having
$J_{J-H} < 12$
show near-IR
excess in the $(Q_{JHK},J_{J-H})$
CMD (Figure \ref{fig-qccd}).

\subsection{The Effectiveness of Classification by Variability Alone}

We see in Table \ref{ori-tbl1} that some discrepancies exist between
the results of the variability-based classification scheme in this
work and those obtained by analysis of optical spectra in 
\citet{bri05b}. Out of the 7 WTTS we find that the
SDSS survey agrees on 4 objects, classing the remaining 3 as non-PMS.
For the 3 confirmed CTTS, we class 2 as WTTS and 1 as non-PMS.

\begin{figure}
\plotone{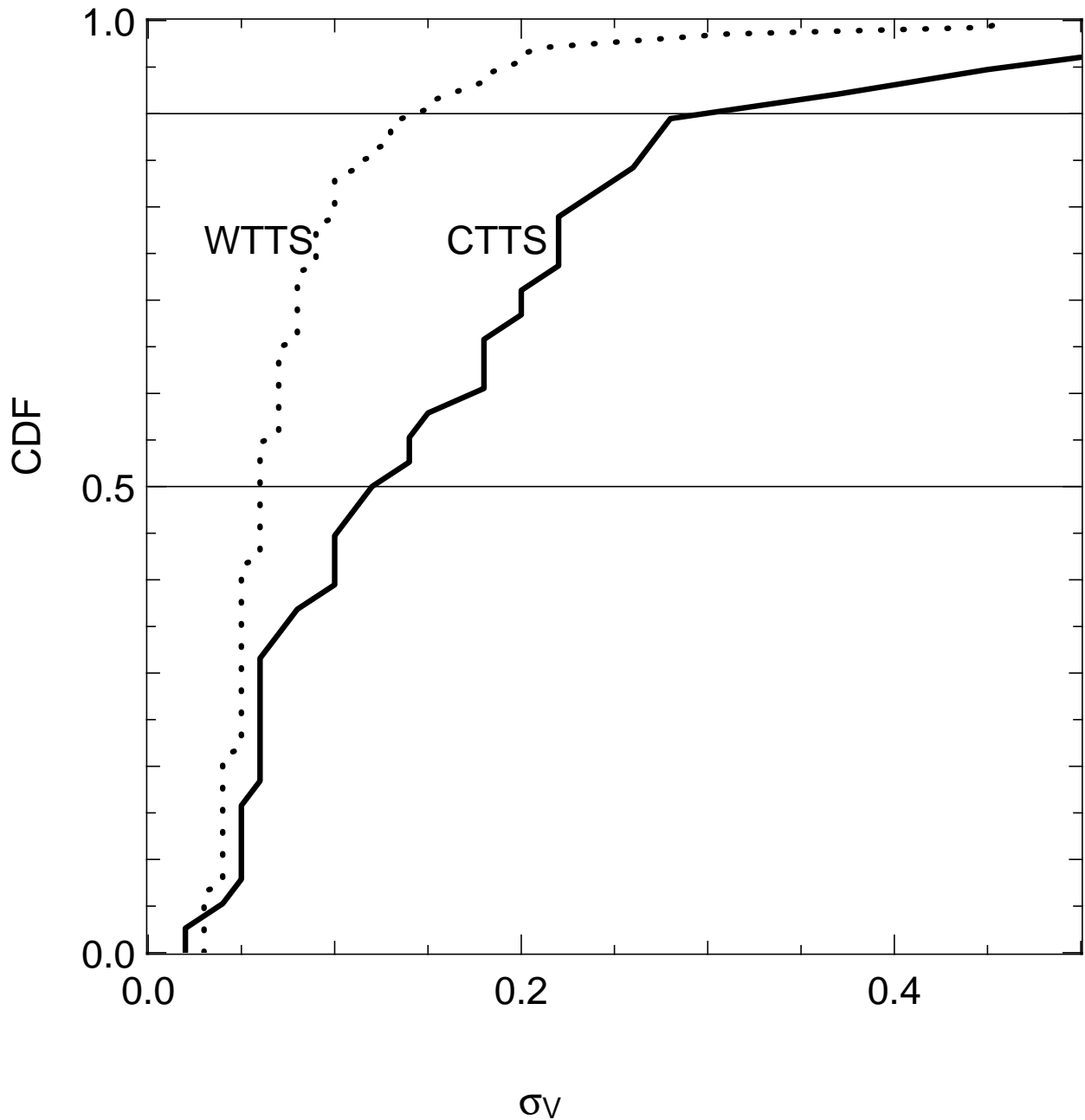}
\caption{{\bf $\sigma_V$ Cumulative Distribution Functions for WTTS and CTTS.}
The Cumulative Distribution Functions (CDF) of $\sigma_V$ are shown
here for the spectroscopically confirmed WTTS ({\it dotted line})
and CTTS ({\it solid line})) from \citet{bri05b} where it is clear
that the CTTS are much more highly variable than the WTTS. 
The median value of $\sigma_V$, at CDF = 0.5, for the WTTS and CTTS 
are 0.06 and 0.12 magnitudes, respectively. The 90th percentile values
of $\sigma_V$ for the WTTS and CTTS have a greater spread of
0.14 magnitudes.
\label{fig-cdf}}
\end{figure}

We use the full sample of \citet{bri05b} to assess the
effectiveness of distinguishing WTTS and CTTS based on a
threshold value of $\sigma_V$. For each 
threshold value of $\sigma_V$ we identify WTTS candidates
as those with $\sigma_V$ below this threshold and CTTS candidates 
those RMS values exceeed this threshold.
In this collection of stars from both the
Orion OB1a and OB1b subassociations there are $N_{WTTS} = 138$
WTTS and $N_{CTTS} = 38$ CTTS, giving a CTTS or inner disk fraction of
21.7\%. We begin by examining the
Cumulative Distribution Function (CDF) for $\sigma_V$ for
both types of PMS stars. In Figure \ref{fig-cdf} we see that
while the CTTS clearly tend to have higher $\sigma_V$ values
than the WTTS, there do exist both high-amplitude WTTS and
low-amplitude CTTS, thus using the amplitude or, in our case, the
RMS of the variability will not be perfectly efficient.

To measure the selection efficiency we consider two selection
metrics, the accuracy and the computed CTTS fraction.
In order to assess the ability to classify individual
stars as WTTS or CTTS we use the accuracy, defined as
\begin{equation}
Accuracy = \frac{N_{TP}}{N_{CTTS}} - \frac{N_{FP}}{N_{WTTS}}
\end{equation}
where $N_{TP}$ and $N_{FP}$ are the number of true positives
and false positives, respectively. We also look at computed
CTTS fraction, given by
\begin{equation}
f_{CTTS} = \frac{N_{TP} + N_{FP}}{N_{CTTS} + N_{WTTS}}.
\end{equation}

\begin{figure}
\plotone{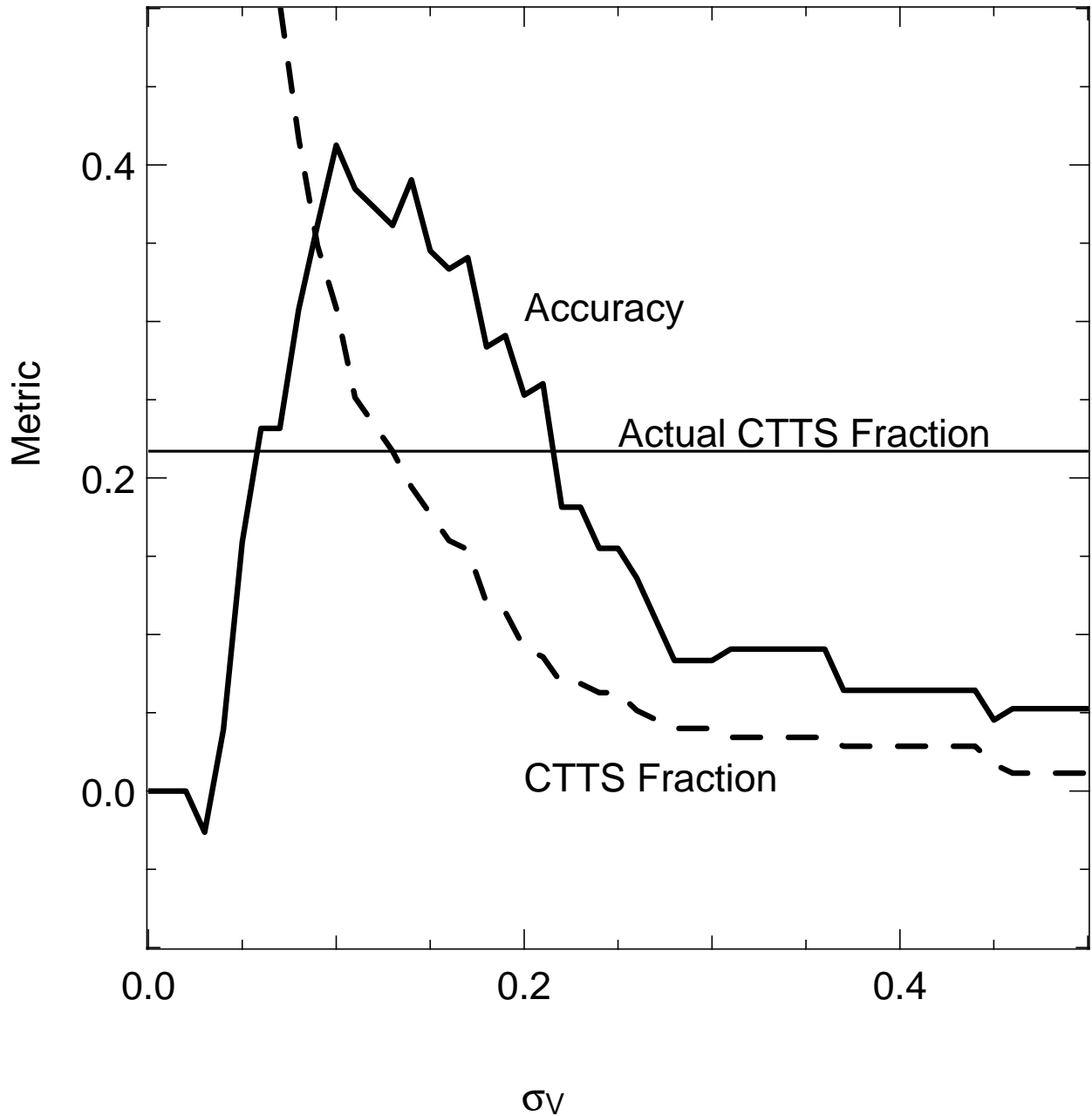}
\caption{{\bf Selection metrics for CTTS classification.}
Two selection metrics for CTTS/WTTS discriminations based on
a threshold $\sigma_V$ value are compared here. To assess the
ability to correct identify individual objects we plot the
accuracy against the threshold ({\it solid line}). The dashed
line shows the computed CTTS fraction ({\it dashed line})
which is compared against
the actual value in the sample ({\it horizontal line}).
Both selection metrics give their best performance for
$0.10 < \sigma_V < 0.15$.
\label{fig-metric}}
\end{figure}

In Figure \ref{fig-metric} we plot both of these selection
metrics against the threshold value of $\sigma_V$. We find that
the peak accuracy of $\sim$0.4 is found for a threshold value
of 0.1 magnitudes and that the computed CTTS fraction is close to the
actual CTTS fraction for $\sigma_V$ thresholds between 0.1 and
0.15 magnitudes. Since the variability amplitude in CTTS 
increases sharply at shorter wavelengths, with typical values
of $\sigma_B/\sigma_V \sim 1.3$ \citep{her94}, we might expect that
the optimal threshold for WTTS/CTTS differentiation using the SDSS $g$
band, which has a similar effective wavelength as $B$, may be in the
range of 0.13 to 0.2 magnitudes.
This suggests that the computed CTTS fraction in our work, which
is based on a $\sigma_g$ threshold of 0.2 magnitudes, may not
be unreasonable.

\section{Results}

\subsection{Distribution}

\begin{figure}
\plotone{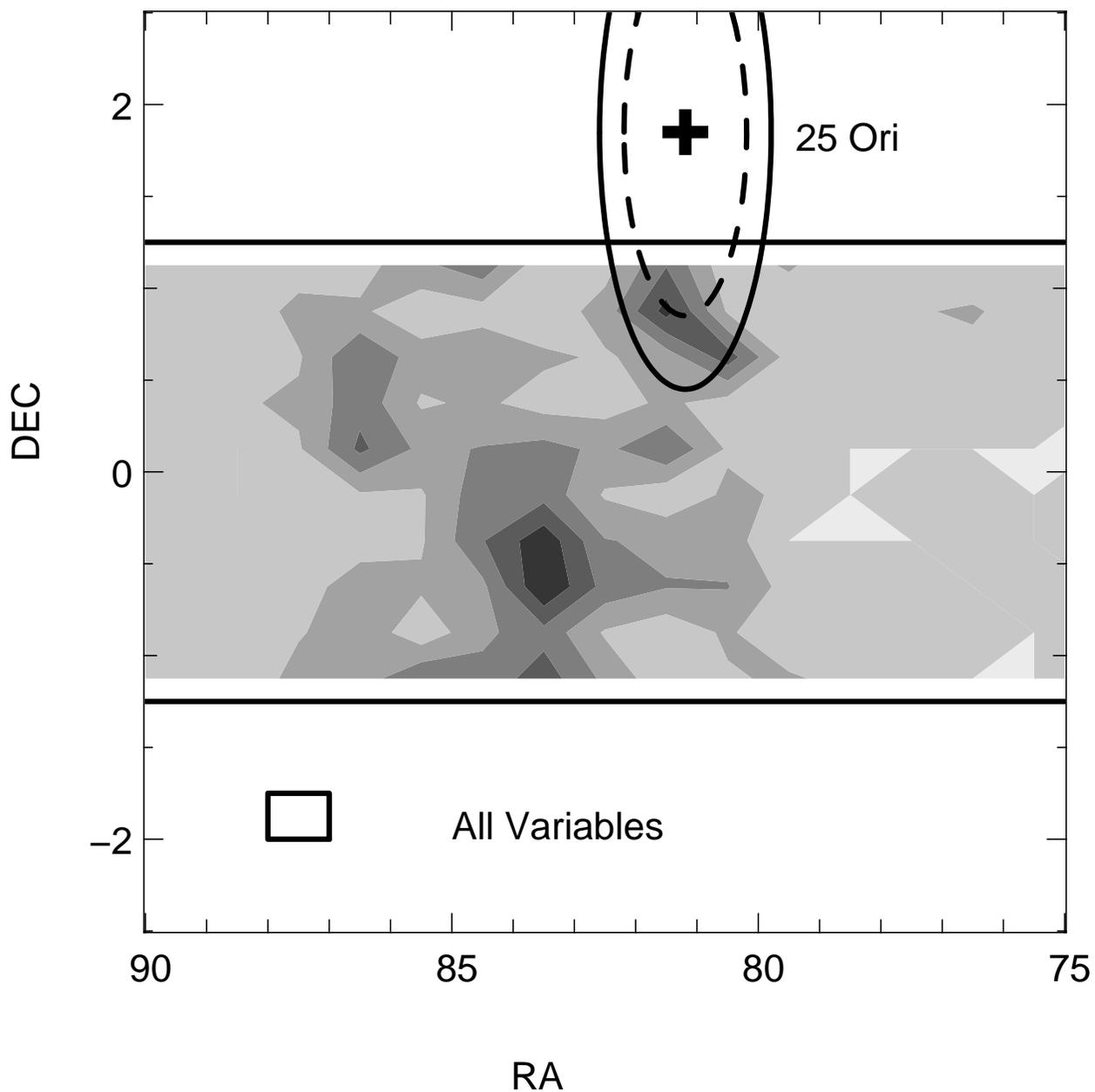}
\caption{{\bf Stellar densities.}
The surface densities of PMS candidates selected by variability
($\sigma_g > 0.05$) are presented here. The
densities are computed based on $1.0\degr$ by $0.25\degr$ regions
(see box in lower left) with the contours uniformly spaced between
0.0 stars deg$^{-2}$  and a maximum of 68 stars deg$^{-2}$.
The L1630 complex ($\alpha_{2000} = 86.5\degr$),
Orion OB1b subassociation, and the 25 Ori group are evident. 
The circles of radii 1.0$\degr$ 
and 1.4$\degr$ are centered
on the Be star 25 Orionis ({\it cross}) and mark the approximate extent of
the 25 Ori group.
\label{fig-radec}}
\end{figure}

The 25 Orionis group is evident in a surface density plot of candidate PMS
stars that are identified on the basis of $\sigma_g > 0.05$. 
In Figure \ref{fig-radec}
there are three obvious groupings of PMS stars which, in order of
increasing R.A., are the 25 Ori group, the Orion OB1b subassociation,
and the NGC 2068/NGC 2071 star formation site in the L1630 cloud.
The surface densities are computed on a $1.0\degr$ by $0.25\degr$ grid
in R.A. and Dec. with the peak value of 68 stars deg$^{-2}$ found at the
center of the Ori OB1b subassociation.

We look for the southern extent of the 25 Ori group by computing the
surface density of the PMS candidates as a function of radial distance 
from the Be star 25 Orionis. Due
to the center of the group being outside of the SDSS imaging area we 
compute the
areas for each radial bin as 
$\int_{r}^{r+{\Delta}r} 2r cos^{-1}(d/r) dr$ 
where the radial bin size ${\Delta}r = 0.4\degr$, the distance
between 25 Ori and the SDSS imaging area $d = 0.6\degr$, and 
$r \ge d$.

\begin{figure}
\plotone{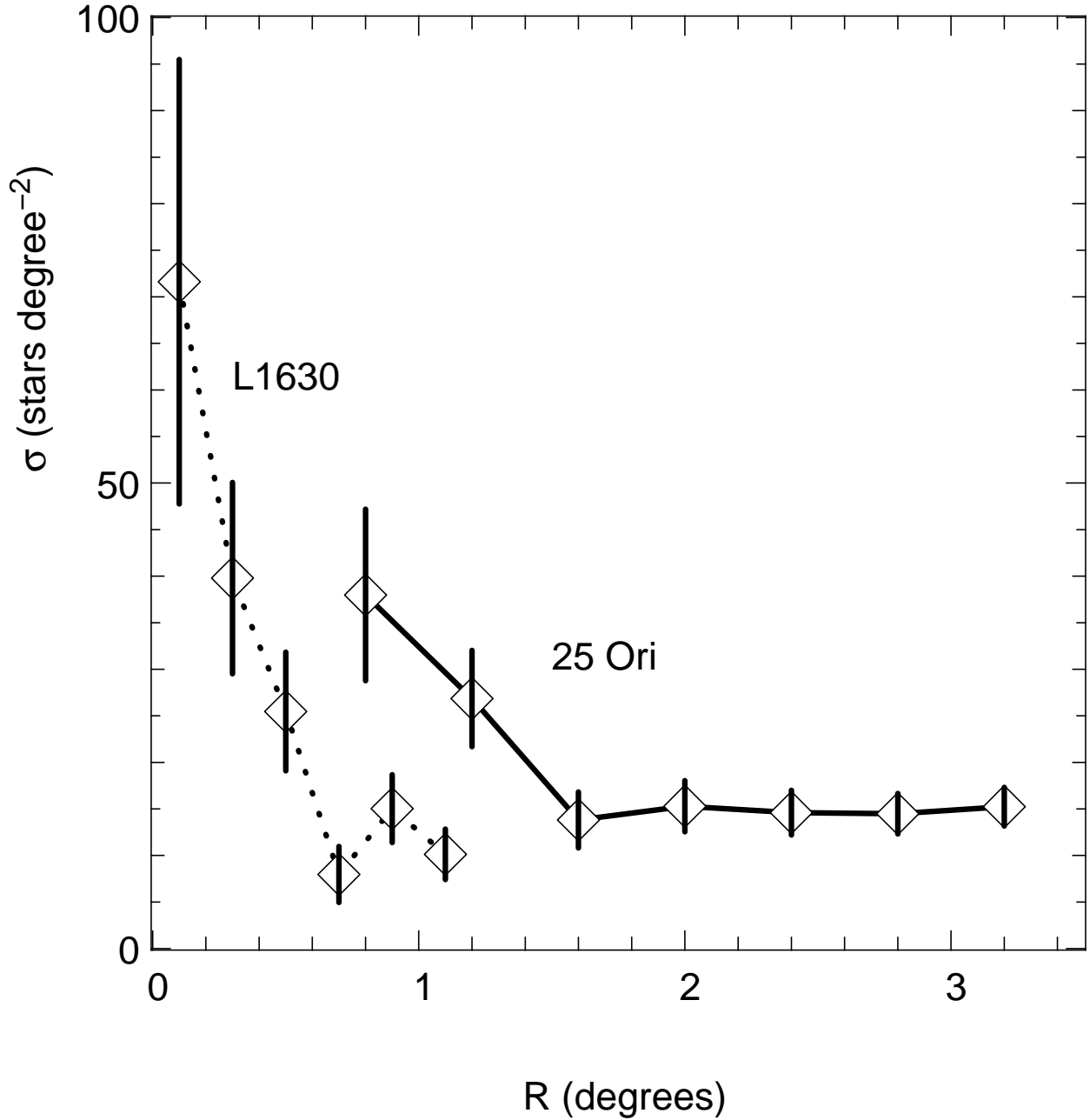}
\caption{{\bf Stellar density versus distance from group center.}
The signatures of the 25 Ori group ({\it solid line}) and the 
NGC 2068/NGC 2071
active star formation site in L1630 ({\it dotted line}) are 
evident in this plot of the
surface density of SDSS PMS candidates using radial bins.
Spacings of 0.4$\degr$ and 0.2$\degr$ are used for the 25 Ori and
L1630 regions, respectively. The innermost bin for the 25 Ori group 
spans distances of 0.6$\degr$ to 1.0$\degr$.
We estimate the outer radius of the
25 Ori group as 1.4$\degr$ and that of the northern L1630
protocluster as 0.6$\degr$. The $\pm 1\sigma$ error bars
are computed based on Poisson statistics. 
\label{fig-density}}
\end{figure}

In Figure
\ref{fig-density} we see that the 25 Ori group blends into the surrounding
dispersed PMS population of Ori OB1a at distances greater than 1.4$\degr$
from the Be star 24 Orionis.
In contrast, 
the NGC 2068/NGC 2071 complex is much more compact and contained within
a radius of 0.6$\degr$, as measured from the center of 
NGC 2071.
The near-IR survey of \citep{lad91} of L1630, complete
to $K = 13$,
showed that the four active star formation sites in L1630 are tightly
clustered with typical effective radii of 0.1$\degr$ or less.
Therefore most of this spatial extent is accounted for by the 
0.3$\degr$ separation between individual NGC 2068 and NGC 2071 protoclusters.
Futhermore, the submillimeter dust continuum study of NGC 2068/NGC 2071 at
450 $\mu$m and 850$\mu$m by \citet{mot01} shows that each of these 
protoclusters consist of filamentary dense cores.

\begin{figure}
\plotone{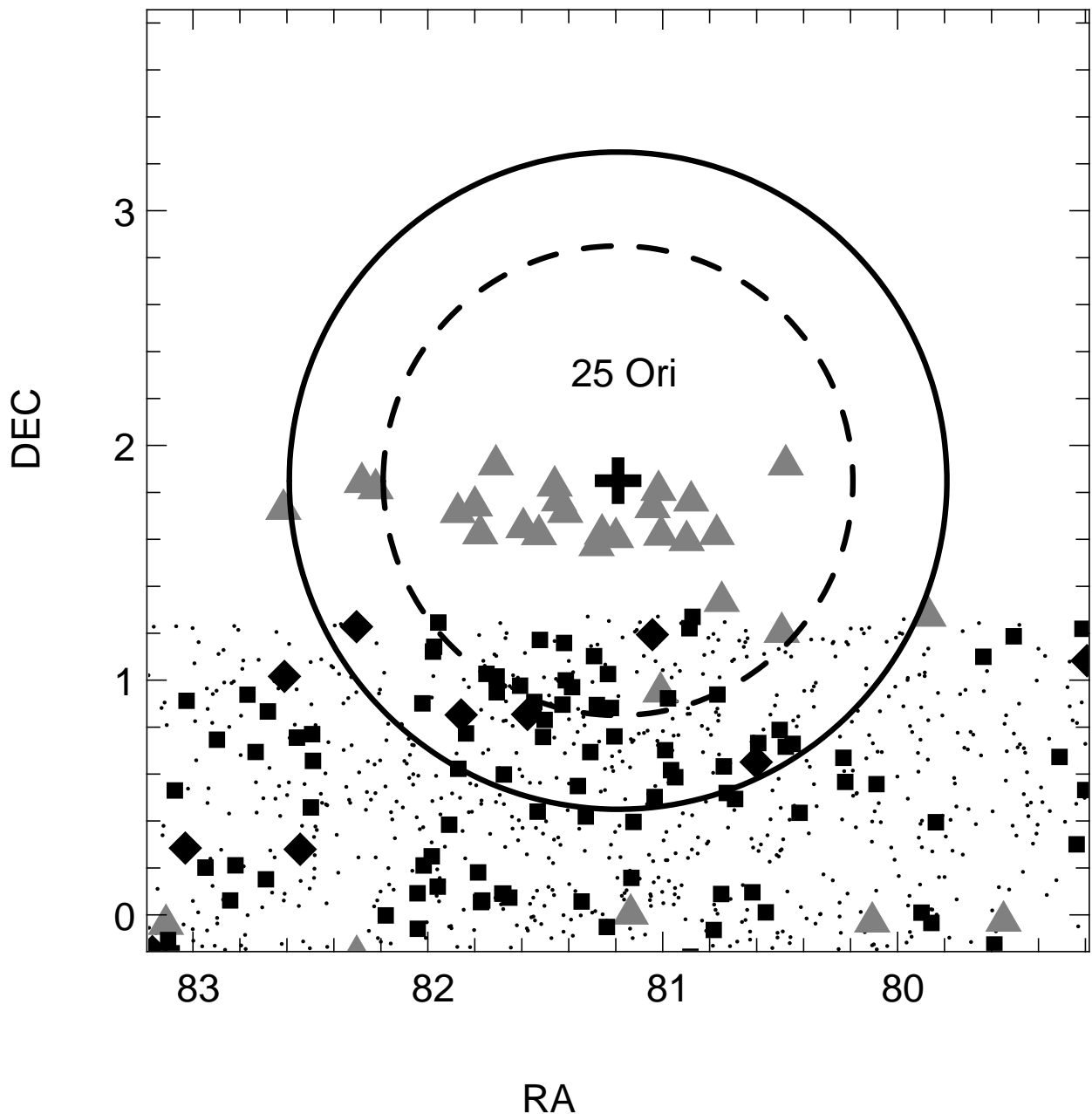}
\caption{{\bf The 25 Ori Region.}
In this figure we show the locations in equatorial coordinates
of the spectroscopically confirmed
T Tauris from the CIDA Variability Survey of Orion ({\it grey triangles}),
SDSS WTTS candidates ({\it small squares}), and SDSS CTTS candidates
({\it diamonds}) within 2 degrees of 25 Ori. The circles of radii 1.0$\degr$ 
and 1.4$\degr$ are centered
on the Be star 25 Ori ({\it cross}) and mark the approximate extent of
the 25 Ori group. The candidate low-mass stars observed for least 2 epochs
by the SDSS are shown as points.
\label{fig-radec_25ori}}
\end{figure}

\begin{figure}
\plotone{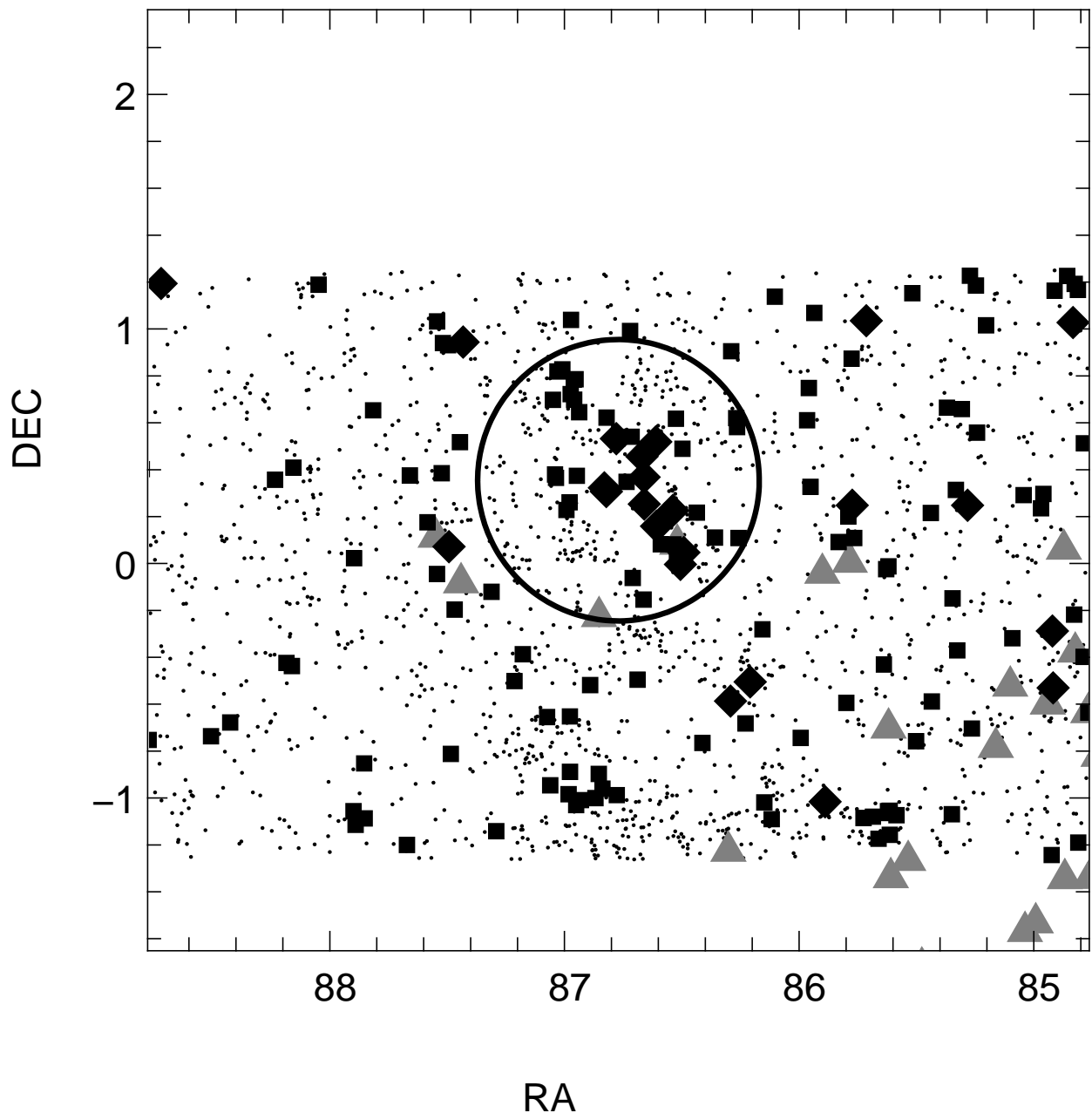}
\caption{{\bf The NGC 2068/NGC 2071 region.}
In this figure we show the locations in equatorial coordinates
of the spectroscopically confirmed
T Tauris from the CIDA Variability Survey of Orion ({\it grey triangles}),
SDSS WTTS candidates ({\it small squares}), and SDSS CTTS candidates
({\it diamonds}) within 2 degrees of NGC 2071. The circle centered on
NGC 2071 has a radius 0.6$\degr$. 
The candidate low-mass stars observed for least 2 epochs
by the SDSS are shown as points.
\label{fig-radec_l1630}}
\end{figure}

In Figure 
\ref{fig-radec_25ori}, which plots the individual stars
within a 4$\degr$ by 4$\degr$ region centered
on 25 Orionis, the clustering of candidate PMS objects around the
central Be star is evident. 
For comparison, Figure \ref{fig-radec_l1630} shows the same sized
region around the NGC 2068/NGC 2071 star formation site. 
In each of these figures we 
identify the SDSS low-mass WTTS and CTTS candidates and the
spectroscopically confirmed T Tauris from the CVSO.

\subsection{CTTS Fraction}

\begin{figure}
\plotone{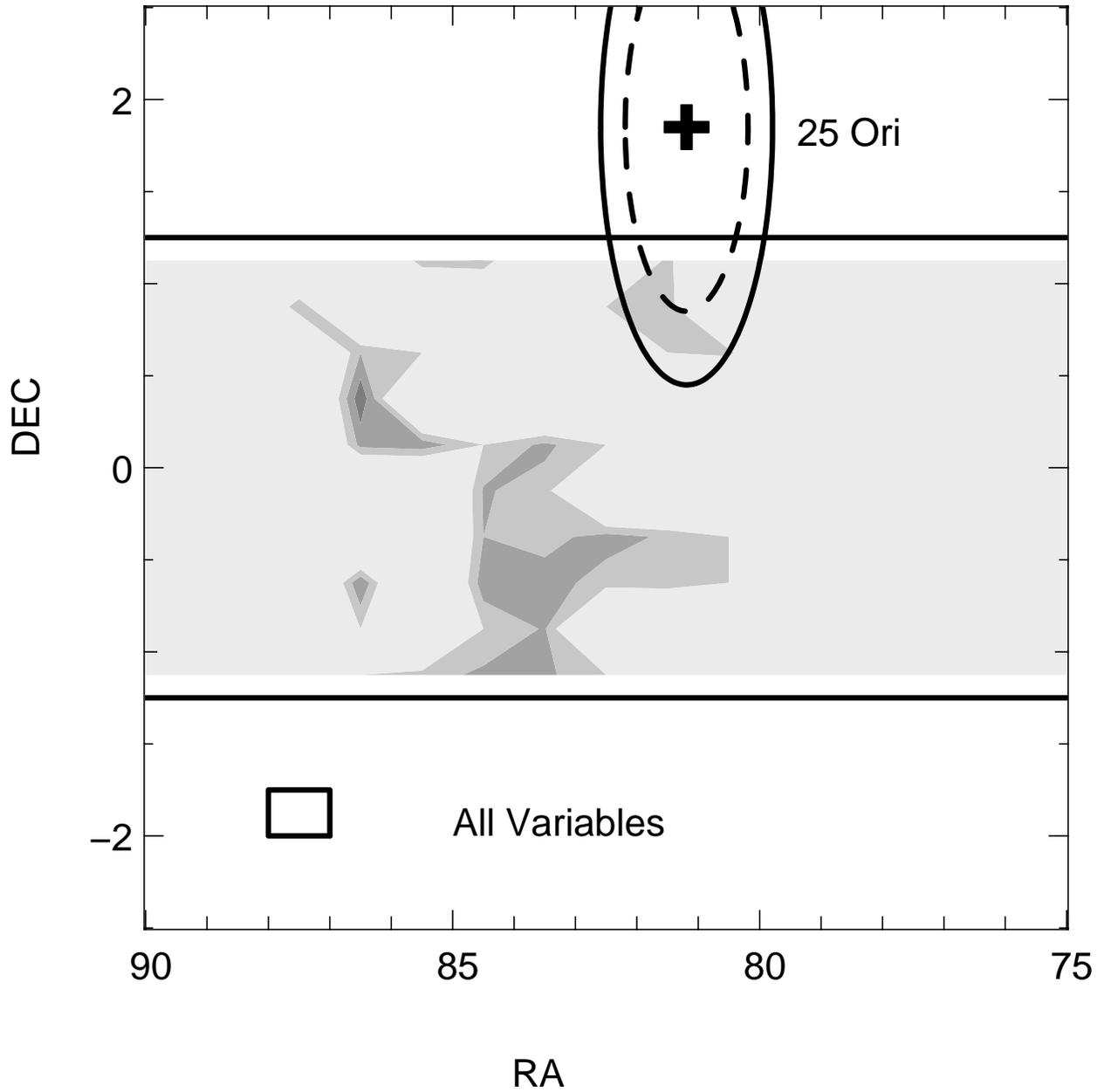}
\caption{{\bf CTTS fractions.}
The fraction of CTTS candidates, 
is shown here using the same samples and gridding as define in
Figure \ref{fig-radec}.
The contours are uniformly spaced between 0.0 and 1.0.
Both the L1630 complex and the Orion OB1b subassociation have relatively
high CTTS fractions.
\label{fig-radecF}}
\end{figure}

We examine the relative fraction of CTTS candidates,
defined as N(CTTS)/(N(CTTS) + N(WTTS)), using the same 
$1.0\degr$ by $0.25\degr$ spatial grid as used by
the surface density map
(Figure \ref{fig-radec}). 
We compute this fraction for each
spatial bin that contains five or more  PMS candidates.
As seen in Figure \ref{fig-radecF} the peak fraction is found in 
the NGC 2068/NGC 2071 complex, which is an active star formation
site. The CTTS fraction is somewhat less in the $\sim$ 2 Myr Ori
OB1b association, and diminished in the 25 Orionis group.

Examination of the CTTS fraction, using the radial bins defined in
Figure \ref{fig-density}, shows that the core of the L1630
site has a CTTS fraction of $0.7\pm0.35$.
This is consistent with the
disk fraction of 0.85$\pm$0.15 found by \citet{hai01}
for another L1630 star formation site, the 0.3 Myr NGC 2024 protocluster. 
In the southern flanking fields
of 25 Ori, however, the CTTS fraction is $\sim 0.1$.
This is
suggestive of an relatively older population than in L1630 or Orion OB1b
since the CTTS fraction decreases with age, due to typical
circumstellar disk lifetimes of order 
1 to 10 Myr \citep{ken95}.

\section{Discussion and Conclusions}

We have mapped the southern extent of the 25 Ori group using multi-epoch
imaging by the Sloan Digital Sky Survey.
By studying the surface densities of PMS candidates in radial bins
centered on the Be star 25 Ori, we see that the low-mass 
population of the 25 Ori group extends out to $\sim 1.4\degr$.
This corresponds to a physical radius of 8 to 11 pc, based
on distances of 330 pc (for Ori OB1a) and 440 pc (for OB1b),
respectively. 

The majority of young ($< 15$ Myr) clusters have radii less
than a few pc \citep{van05}. While there exist young clusters that
have significantly larger physical dimensions these are probably
not gravitationally bound.
The large spatial extent of the 25 Ori group suggests that it is an
unbound association, rather than a open cluster, although
this is a preliminary classification based only on the southern
flanking fields. 

Comparison of variability and near-IR excess have led to the adoption
of  selection criteria for CTTS candidates based on multi-band
variability that are $\sigma_g > 0.2$ and $\sigma_z > 0.05$, with the
$g$ band criterion a factor of four times more restrictive than used in
\citet{mcg05}.
By classifying objects on the
basis of multi-band variability as probable WTTS or CTTS candidates we 
detect the known star formation complexes in the Orion equatorial
region (L1630 and Orion OB1b) by virtue of their high CTTS fraction, with the
core of the NGC 2068/NGC 2071 complex in the northern portion of
the L1630 cloud having a CTTS fraction $\sim 0.7$.

We find that both the 25 Ori group and the Orion OB1a 
subassociation have CTTS fractions of $\sim$0.1, similar to that
found by \citet{bri05b}. In Figure 12. of 
\citet{bri05b} we that this CTTS, or inner disk, fraction,
is consistent with ages between 5 Myr and 30 Myr.

Future work involving this group includes mapping of the central
region and detailed comparison of its structure with a number
of young clusters and associations. Our goal is to determine if a
gravitationally bound core to the 25 Ori group exists that may
evolve into an open cluster.

\acknowledgements

We thank the anonymous referee for comments that have 
improved this paper.
P.M.M. acknowledges support from LANL Laboratory Directed
Research and Development (LDRD) program 20060495ER.
Funding for the SDSS and SDSS-II has been provided by the Alfred
P. Sloan Foundation, the Participating Institutions, the National
Science Foundation, the U.S. Department of Energy, the National
Aeronautics and Space Administration, the Japanese Monbukagakusho, the
Max Planck Society, and the Higher Education Funding Council for
England.  The SDSS Web Site is http://www.sdss.org/.

The SDSS is managed by the Astrophysical Research Consortium for
the Participating Institutions. The Participating Institutions are the
American Museum of Natural History, Astrophysical Institute Potsdam,
University of Basel, Cambridge University, Case Western Reserve
University, University of Chicago, Drexel University, Fermilab, the
Institute for Advanced Study, the Japan Participation Group, Johns
Hopkins University, the Joint Institute for Nuclear Astrophysics, the
Kavli Institute for Particle Astrophysics and Cosmology, the Korean
Scientist Group, the Chinese Academy of Sciences (LAMOST), Los Alamos
National Laboratory, the Max-Planck-Institute for Astronomy (MPA), the
Max-Planck-Institute for Astrophysics (MPIA), New Mexico State
University, Ohio State University, University of Pittsburgh,
University of Portsmouth, Princeton University, the United States
Naval Observatory, and the University of Washington.

This publication makes use of data products of the Two Micron
All Sky Survey, which is a joint project of the University of
Massachusetts and the Infrared Processing and Analysis Center/California
Institute of Technology, funded by the National Aeronautics and Space
Administration and the National Science Foundation.

Facilities: \facility{SDSS}, \facility{2MASS}

%\clearpage

\begin{figure}
\plotone{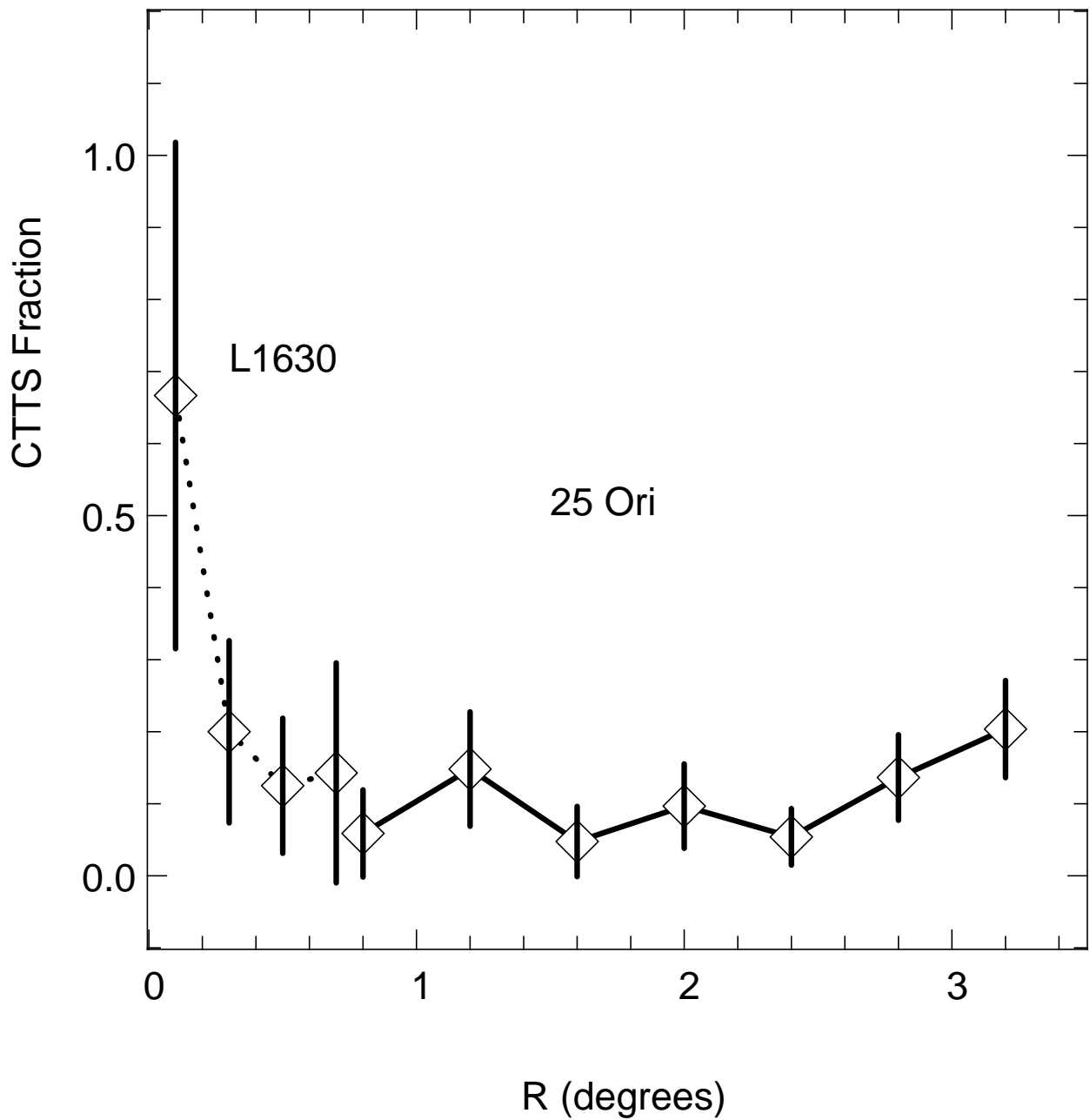}
\caption{{\bf CTTS fraction versus distance from group center.}
The 25 Ori group has a mean CTTS fraction ({\it solid line})
 as $\sim 0.1$, as seen in
this plot of CTTS fraction against radial distance from 25 Orionis.
This fraction is identical to that observed in the Orion OB1a
field surrounding the 25 Ori group. 
In contrast, the central region of NGC 2068/NGC 2071 protocluster
({\it dotted line}) has a high CTTS fraction of $\sim$0.7.
The CTTS fractions are computed using the same radial bins as in
Figure \ref{fig-density}.
The $\pm 1\sigma$ error bars
are computed based on Poisson statistics. 
\label{fig-densityz}}
\end{figure}

\clearpage
\oddsidemargin=-1cm
\tabletypesize{\scriptsize}
\begin{deluxetable}{lllllllll}
\tablecaption{Matches between the SDSS and CVSO Surveys\label{ori-tbl1}}
\tablehead{
\colhead{CVSO} &
\colhead{SDSS Name} &
\colhead{$g$} &
\colhead{$r$} &
\colhead{$i$} &
\colhead{$z$} &
\colhead{R\tablenotemark{a}} &
\colhead{Class\tablenotemark{b}} &
\colhead{Confirmed Class\tablenotemark{c}}
}
\startdata
CVSO   1 & SDSS J051058.78-000803.1 & 17.96 & 16.58 & 15.12 & 14.33 &  3.98 & W & C \\
CVSO  13 & SDSS J052242.82-005306.5 & 17.17 & 15.71 & 14.70 & 14.07 &  2.78 & W & W \\
CVSO  49 & SDSS J052806.91-001806.0 & 18.47 & 17.06 & 15.74 & 14.98 &  2.31 & W & W \\
CVSO  81 & SDSS J053055.62-005540.0 & 17.02 & 15.69 & 14.51 & 13.74 &  3.18 & - & W \\
CVSO 124 & SDSS J053349.04-004625.9 & 16.81 & 15.44 & 14.29 & 13.63 &  3.47 & W & W \\
CVSO 142 & SDSS J053530.03-002822.7 & 17.04 & 15.60 & 14.48 & 13.93 &  3.55 & - & W \\
CVSO 156 & SDSS J053747.02-002007.2 & 16.73 & 15.38 & 14.25 & 13.50 &  3.92 & - & C \\
CVSO 162 & SDSS J053830.84-011024.3 & 16.74 & 15.36 & 14.51 & 14.05 &  4.58 & - & W \\
CVSO 164 & SDSS J053853.84-004920.0 & 17.51 & 16.14 & 14.64 & 13.56 &  4.43 & W & C \\
CVSO 168 & SDSS J053904.29-003810.0 & 18.19 & 16.76 & 15.34 & 14.48 &  4.36 & W & W \\
\enddata
\tablenotetext{a}{Distance in degrees from the Be star 25 Orionis.}
\tablenotetext{b}{Preliminary classification as not PMS (-), WTTS (W) or 
CTTS (C)
based on multi-band variability.}
\tablenotetext{c}{Classification as WTTS (W) or CTTS (C) on the
basis of optical spectroscopy \citep{bri05b}.}
\end{deluxetable}

%\tabletypesize{\scriptsize}
%\begin{deluxetable}{lrrllllll}
%\tablecaption{Candidate Low-Mass PMS in the Orion Equatorial Region\label{ori-tbl2}}
%\tablehead{
%\colhead{Name} &
%\colhead{RA} &
%\colhead{DEC} &
%\colhead{$g$} &
%\colhead{$r$} &
%\colhead{$i$} &
%\colhead{$z$} &
%\colhead{R\tablenotemark{a}} &
%\colhead{Class\tablenotemark{b}}
%}
%\startdata
%SDSS J050013.47+011349.1 &   75.0562 &    1.2303 & 19.90 & 18.24 & 16.67 & 15.77 &  6.17 & W \\
%SDSS J050013.86+005807.4 &   75.0578 &    0.9687 & 16.85 & 15.44 & 14.57 & 14.09 &  6.20 & W \\
%SDSS J050058.65+002106.9 &   75.2444 &    0.3519 & 19.92 & 18.29 & 16.14 & 15.03 &  6.13 & W \\
%SDSS J050333.53-004138.2 &   75.8897 &   -0.6940 & 18.79 & 17.40 & 15.84 & 14.96 &  5.88 & C \\
%SDSS J050531.61+002616.3 &   76.3817 &    0.4379 & 20.56 & 18.78 & 16.82 & 15.88 &   5.01 & W \\
%\enddata
%\tablecomments{Table \ref{ori-tbl2} is presented in its entirety in the
%electronic edition of the Astronomical Journal. A portion is shown here 
%for guidance regarding its form and content.}
%\tablenotetext{a}{Distance in degrees from the Be star 25 Orionis.}
%\tablenotetext{b}{Preliminary classification as WTTS (W) or CTTS (C)
%based on multi-band variability.}
%\end{deluxetable}

\oddsidemargin=-1cm
\tabletypesize{\scriptsize}
\begin{deluxetable}{lrrllllll}
\tablecaption{Candidate Low-Mass PMS in the Orion Equatorial Region\label{ori-tbl2}}
\tablehead{
\colhead{Name} &
\colhead{RA} &
\colhead{DEC} &
\colhead{$g$} &
\colhead{$r$} &
\colhead{$i$} &
\colhead{$z$} &
\colhead{R\tablenotemark{b}} &
\colhead{Class\tablenotemark{c}}
}
\startdata
SDSS J050013.47+011349.1 &   75.0562 &    1.2303 & 19.90 & 18.24 & 16.67 & 15.77 &  6.17 & W \\
SDSS J050013.86+005807.4 &   75.0578 &    0.9687 & 16.85 & 15.44 & 14.57 & 14.09 &  6.20 & W \\
SDSS J050058.65+002106.9 &   75.2444 &    0.3519 & 19.92 & 18.29 & 16.14 & 15.03 &  6.13 & W \\
SDSS J050333.53-004138.2 &   75.8897 &   -0.6940 & 18.79 & 17.40 & 15.84 & 14.96 &  5.88 & W \\
SDSS J050531.61+002616.3 &   76.3817 &    0.4379 & 20.56 & 18.78 & 16.82 & 15.88 &  5.01 & W \\
SDSS J050548.68-000332.8 &   76.4528 &   -0.0591 & 19.06 & 17.33 & 16.35 & 15.86 &  5.11 & W \\
SDSS J050549.25+005120.9 &   76.4552 &    0.8558 & 19.22 & 17.54 & 16.50 & 15.94 &  4.84 & W \\
SDSS J050601.35-004831.0 &   76.5057 &   -0.8086 & 18.14 & 16.56 & 15.93 & 15.64 &  5.39 & W \\
\enddata
\tablenotetext{a}{Distance in degrees from the Be star 25 Orionis.}
\tablenotetext{b}{Preliminary classification as WTTS (W) or CTTS (C)
based on multi-band variability.}
\end{deluxetable}

\end{document}